\documentclass[%
 reprint,
superscriptaddress,
floatfix,
 amsmath,amssymb,
 aps,
 pra,
]{revtex4-1}

\usepackage{graphicx}%
\usepackage[displaymath, mathlines,running,columnwise]{lineno}
\usepackage{dcolumn}%
\usepackage{bm}%
\usepackage[pdftex]{hyperref}%
\usepackage{ bbold }
\usepackage{ dsfont }
\usepackage{amsthm} %
\usepackage{braket}
\usepackage{natbib}
\usepackage{float}
\usepackage{multirow}
\usepackage{xcolor}

\usepackage{fancyhdr}
\newcommand{\tr}{\text{tr}}
\newcommand{\one}{\mathbb{1}}

\def\lb{\left(}
\def\rb{\right)}

\newcommand{\R}{\mathds{R}}
\newcommand{\ketbra}[2]{|#1\rangle\langle #2|}

\newtheorem{theorem}{Theorem}

\newtheorem{lem}{Lemma}

\newtheorem{example}{Example}
\let\olddefinition\example
\renewcommand{\example}{\olddefinition\normalfont}
\newtheorem{definition}{Definition}
\let\olddefinition\definition
\renewcommand{\definition}{\olddefinition\normalfont}
\begin{document}

\title{Noise-robust exploration of many-body quantum states on near-term quantum devices}%
 \author{Johannes Borregaard}%
 
 \affiliation{%
QMATH, Department of Mathematical Sciences, University of Copenhagen, Universitetsparken 5, 2100 Copenhagen, Denmark}%

\affiliation{%
Qutech and Kavli Institute of Nanoscience, Delft University of Technology, 2628 CJ Delft, The Netherlands}
 \author{Matthias Christandl}
\author{Daniel Stilck Fran\c{c}a}

 \email{dsfranca@math.ku.dk}
\affiliation{%
QMATH, Department of Mathematical Sciences, University of Copenhagen, Universitetsparken 5, 2100 Copenhagen, Denmark}%

\date{\today}%

\begin{abstract}
We describe a resource-efficient approach to studying many-body quantum states on noisy, intermediate-scale quantum devices. We employ a sequential generation model that allows us to bound the range of correlations in the resulting many-body quantum states. From this, we characterize situations where the estimation of local observables does not require the preparation of the entire state. Instead smaller patches of the state can be generated from which the observables can be estimated. This can potentially reduce circuit size and number of qubits for the computation of physical properties of the states. Moreover, we show that the effect of noise decreases along the computation.
Our results apply to a broad class of widely studied tensor network states and can be directly applied to near-term implementations of variational quantum algorithms. 
\end{abstract}

\pacs{Valid PACS appear here}%
\maketitle

\section{Introduction}
\normalfont
Quantum computers offer computational power fundamentally different from classical computers. A universal quantum computer may solve classically intractable problems within areas ranging from many-body physics to quantum chemistry~\cite{Georgescu2014}. There has been impressive experimental progress in developing quantum computers on different architectures~\cite{Neill2018,Zhang2017,Bernien2017}. Although achieving fault-tolerant computation remains a challenge, noisy intermediate-scale quantum (NISQ) devices are expected to be available in the near future~\cite{Preskill2018}. These are devices containing a few hundred qubits with small error rates but without error-correction. An outstanding question is what kind of computations such devices may facilitate. 

Algorithms designed for NISQ devices should run on a moderate number of qubits and be resilient to noise. The specific hardware may also pose further restrictions regarding the connectivity of the device, as not all qubits can interact directly with each other~\cite{Neill2018,Bernien2017}.  Promising frameworks that fulfill these conditions are the quantum approximate optimization algorithm (QAOA)~\cite{Farhi2014} and the quantum variational eigensolver (VQE)~\cite{Peruzzo2014,McClean2016}. In these frameworks, the task of the quantum computer is roughly speaking to compute the expectation value of local Hamiltonians on some many-body quantum state. Recent work has characterized a number of conditions for which this can be done in a noise-robust way~\cite{Temme2017,Sharma2019,Murali2019,kim2017noise,kimswingle}. Due to the limited resources of NISQ devices, it is also important to run such algorithms as efficiently as possible in terms of circuit size and number of qubits. 

We address this outstanding problem by developing a general framework for computing physical properties of quantum many-body states efficiently on NISQ devices. In particular, we upper bound the circuit size and number of qubits necessary to estimate the expectation values of local observables. Importantly, these bounds can significantly decrease the resource requirements compared to previous works for a number of circuit topologies and sizes. Specifically, we are able to show that the energy of a many-body quantum state can be estimated with a constant-sized quantum circuit if the correlation functions exhibit an exponential decay. This is the case for non-trivial states such as ground states of gapped Hamiltonians, surface codes and quantum states described by a multi-scale entanglement renormalization ansatz (MERA) or the larger class of deep MERA (DMERA)~\cite{kim2017noise,kimswingle}. The latter is believed to capture Chern insulators.

Our framework is akin to sequentially generated~\cite{Schon2005,Banuls2008,Gross2007} or finitely correlated states~\cite{Fannes1992}. This enables us to control the size of the past causal cone~\cite{Giovannetti2008,Evenbly2009,shebab} of local observables. Combined with the notion of mixing rate of local observables under the circuit~\cite{kim2017noise,kimswingle} we determine after how many layers of the circuit, expectation values stabilize. To estimate these expectation values, it suffices to implement the potentially small subset of the circuit under which they stabilize instead of producing the entire many-body state or its past causal cone. Consequently, the necessary number of qubits and quantum gates can be reduced significantly from scaling with the size of the many-body state to even a constant number.

\begin{figure}[h!]
 \centering
\includegraphics[width=0.9\columnwidth,trim={0cm 5cm 4.5cm 1.5cm},clip]{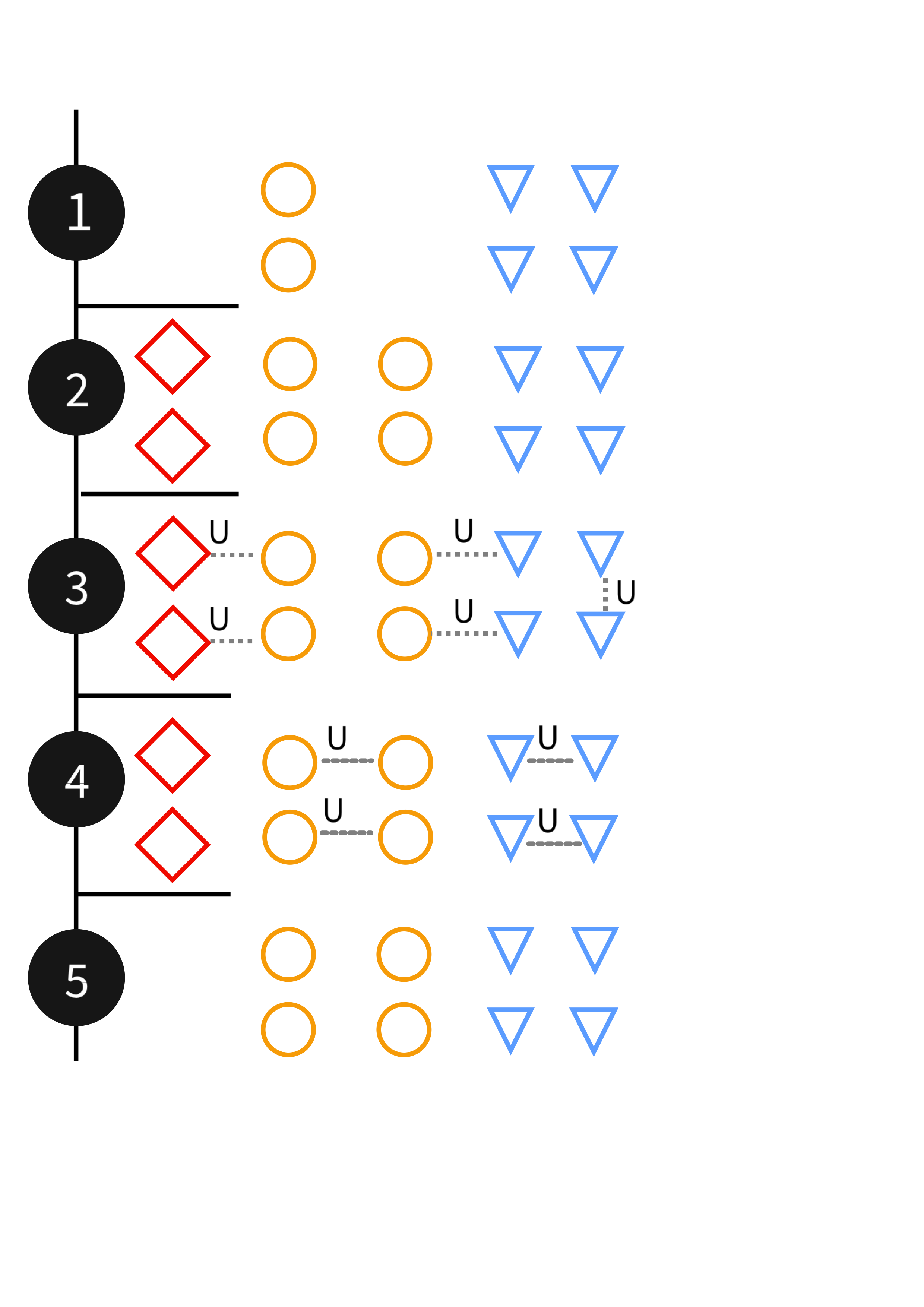}

\caption{(Generation procedure): Example of one iteration of the generation procedure broken down into five steps. The first line (1) represents the initial system before the iteration. We start with two system qubits (orange circles) and a bath (blue triangles). The first operation (line 1 to 2) is to add two new system qubits and two auxiliary qubit (red diamond). The new qubits are placed to the right of the old ones. The second operation (line 2 to 3) is to act with a unitary $\mathcal{U}$ between the indicated qubits and from (3) to (4) we apply another layer of unitaries and, thus, $D=2$ for this example. Finally, in line (4) to (5) we discard the auxiliary systems.\label{fig:interactionscheme}}

\end{figure}

\section{Results}
\subsection{Basic setup}
We consider three basic operations, which are iterated $T$ times to generate a many-body quantum state. The first operation adds qubits to the existing system. The second operation lets them interact with each other and a bath via a constant depth circuit. In the third operation, some of the existing qubits may be discarded. By introducing a separate bath, we allow for situations where a fixed sized quantum processor (the bath) iteratively prepares a quantum state on the system qubits. This allows e.g. the construction of arbitrary matrix product states (see Example 1 below).

More specifically, we will start with a system $S_0$ consisting of $n_0$ qubits initialized in some fixed state $\rho_0$ and a bath system $B$ consisting of $s_B$ qubits initialized in some fixed state $\rho_B$. At each iteration $t$, we introduce new subsystems $S_t$ with $n_t$ qubits and ancillary states $A_t$ with $a_t$ qubits, all initialized in some fixed quantum state. These new subsystems then interact with the existing ones and finally, the ancillary system is discarded, which concludes the iteration. The procedure is iterated for a total of $T$ iterations to produce the entire state.  

\subsection{Interaction scheme} 
The structure of the final quantum state is determined by the allowed interactions between qubits during the iterative preparation. In order to get a handle on how the correlations of the final state evolve during the preparation, we will fix the allowed interactions according to a given interaction scheme. Our construction of such an interaction scheme is inspired by so-called graph states~\cite{Raussendorf2003}.  These can be visualised by letting the vertices of a graph denote qubits and edges denoting correlations between them. In a similar way, we define an interaction scheme as a sequence of $T$ graphs $\{G_t=(V_t,E_t)\}$ where a vertex ($V_t$) denotes a qubit and an edge ($E_t$) implies that it is possible to implement unitary gates between these two qubits in that time step. We further restrict this unitary gate such that at most $D$ rounds of two qubit operations are applied with the condition that on each round at most one unitary acts on each qubit. Fig.~\ref{fig:interactionscheme} showcases a simple example of one iteration of the procedure illustrating all aspects of the framework. While such a simple interaction scheme serves the purpose of illustration, our framework encompasses arbitrary interaction graphs in each iteration step. This allows it to capture a number of widely studied tensor networks states, as illustrated in the two examples below. 

\begin{example}[Matrix product states and higher dimensional versions]
Matrix product states (MPS) have been studied in a sequential interaction picture~\cite{Perez-Garcia2006,Schon2007} and adapt naturally to our framework. The initial system $S_0$ consists of one qubit and the dimension of the bath gives the bond dimension. At each iteration $t$, we add a system consisting of one qubit $S_t$ to the system. The graph $G_t$ only has edges between the qubits in the bath and the newest added qubit $S_t$. Note that we are considering a proper subset of MPS since we restrict to unitaries implementable with $D$ two-qubit gates for each edge.  
Nevertheless, the bond dimension of the resulting MPS scales exponentially with the number of qubits in the bath.
It is straightforward to generalize such sequentially generated states by considering the case in which a bath interacts with a subsystem of dimension $d$ rather than a single qubit at each iteration~\cite{kim2017noise,Banuls2008}. 
\end{example}
\begin{example}[Deep multiscale entanglement renormalization ansatz]
The DMERA, introduced in~\cite{kimswingle}, is a variation of the MERA~\cite{Evenbly2009} tailored for NISQ devices. In our framework, the initial system $S_0$ consists of one qubit and there is no bath. We then define the graphs $G_t$ recursively: at each iteration we add one qubit in between every existing qubit and nearest neighbors interact, resulting in a tree structure. 
\end{example}  

\subsection{Past causal cone}
Formally, the final state of the system can always be written as 
\begin{linenomath}
\begin{align}\label{equ:definitionstate}
\rho=\tr_B\left[ \lb\Phi_{[0,T]}\rb \lb\rho_0\otimes \rho_B\rb\right],
\end{align}
\end{linenomath}
where $\Phi_{[0,T]}=\Phi_T\circ\Phi_{T-1}\circ\ldots\circ\Phi_0$ and $\Phi_t$ are quantum channels of the form $\Phi_t=\mathcal{D}_t\circ \mathcal{U}_t\circ \mathcal{A}_t$.  Here $\mathcal{A}_t$ adds the new subsystems and auxiliary qubits, $\mathcal{U}_t$ is a unitary channel that consists of $D$ two-qubit gates for each edge in $G_t$ and $\mathcal{D}_t$ traces out the ancillary systems.
An important property of our framework is that it allows to bound the number of qubits that can influence the value of a local observable, referred to as the causal cone of an observable~\cite{Giovannetti2008,Evenbly2009}. The growth of the casual cone depends on the geometry of the graph $G_t$. 
To see this, consider an observable $O_T$ on the final state $\rho$. According to Eq.~(\ref{equ:definitionstate}), the expectation value is
\begin{linenomath}
\begin{align*}
\tr\lb\rho O_T\rb=\tr\lb\left[ \Phi_{[0,T]}^* \lb O_T\otimes\one_B\rb\right]\rho_0\otimes \rho_B\rb
\end{align*}
\end{linenomath}
where $\Phi^*_{[t,T]}=\Phi_{t}^*\circ \Phi_{t+1}^{*}\circ\cdots\circ \Phi_{T}^*$. Here $\Phi_t^{*}$ is the evolution in the Heisenberg picture. 

We can use our framework to bound the size of the radius of the support of $O_T$ on the final state i.e. the number of qubits on the final state that $O_T$ involves. Going back to the $t$'th iteration, we denote the observable $O_{t}=\Phi^*_{[t,T]} \lb O_T\otimes\one_B\rb$. Let $R(O_{t})$ be the radius of the smallest ball in $G_{t}$ containing the support of $O_{t}$. That is, $O_{t}$ differs from the identity on qubits that are at most $2R(O_t)$ edges away in the graph $G_{t}$. To analyse the growth of the support and its past causal cone we
consider the action of $\Phi_T^*= \mathcal{A}^*_T\circ\mathcal{U}^*_T\circ\mathcal{D}^*_T$. First, $\mathcal{D}^*_T$ acts by tensoring the identity operator on the auxiliary qubits, not increasing the support. In the next step, $\mathcal{U}_T^*$ increases the support.
As $O_T$ has radius $R(O_T)$, it will be mapped to an observable with radius at most $R(O_T)+D$ by $\mathcal{U}_T^*$ according to the locality assumptions of $\mathcal{U}_T^*$ (i.e.  the restriction of $D$ two-qubit gates for each edge). The map $\mathcal{A}_T^*$ will then map this observable to $O_{T-1}$ supported on qubits that correspond to vertices in $G_{T-1}$, as it traces out all the qubits added at iteration $T$. This can potentially decrease the support of the observable, as in DMERA. 

Given the graphs $G_t$ and a constant $D$, it is straightforward to track the support of the observable and the past causal cone with the above procedure. This allows us to find the maximum number of unitaries ($N_U(t,r)$) and qubits ($N_Q(t,r)$) in the past causal cone of an observable with radius of support of $r$ on the final state going back to iteration $t$ . Note that $N_Q(t,r)$ keeps track of the total number of qubits necessary to implement the past causal cone and thus also includes those that were discarded at a previous step. 

\subsection{Estimating local observables}
So far we have devised a way of keeping track of the unitaries in the past causal cone of local observables. However, we are  also interested in quantifying how much each iteration of the past causal cone contributes to the expectation value. In case the expectation value of the observable stabilizes after a couple of iterations, we can find smaller quantum circuits than the entire causal cone that will approximate the desired expectation value.

Inspired by Refs.~\cite{kim2017noise,kimswingle}, we assume that the maps $\Phi^*_{[t,T]}$ are locally mixing. To this end, let us define the mixing rate as:
\begin{equation*}
\delta(t,r)\equiv \sup\limits_{R(O_T)\leq r,\|O_T\|_\infty\leq1}\inf_{c\in \R}\|\Phi^*_{[t,T]}\lb O_T\rb-c\one\|_\infty.
\end{equation*}
Here $\|\cdot\|_\infty$ is the operator norm. The mixing rate, $\delta(t,r)$, quantifies how close observables on the final state, whose support is contained in a ball of radius $r$, are to the identity after going back to the $t$'th iteration of the evolution in the Heisenberg picture. 
Intuitively speaking, $\delta(t,r)$  measures how many steps of the circuit contribute to the expectation value of local observables before it stabilizes, as we are interested in the regime in which $c$ approaches $\tr\left(\rho O\right)$ for large enough $t$.
This is also connected to the memory of the evolution~\cite{Kretschmann2005}. The next lemma formalizes this intuition (see Methods for a proof):
\begin{lem}\label{lem:mixingcorrel}
\normalfont
Let $O_T$ be an observable supported in a ball of radius $r$. Then
\begin{linenomath}
\begin{align}\label{equ:mixingandapprox1}
\left| \tr\lb \Phi_{[t,T]}\lb \rho'\rb O_T\rb-\tr\lb \rho O_T\rb\right|\leq 2\delta(t,r)\|O_T\|_\infty,
\end{align}
\end{linenomath}
where $\rho=\tr_B\left[\Phi_{[0,T]}\lb\rho_0\otimes\rho_B\rb\right]$, which holds for all $\rho'$.
\end{lem} 
In other words, only the last $T-t$ steps of the circuit are necessary to approximately compute the expectation value of $O_T$ up to an error of $2\delta(t,r)\|O_T\|_\infty$. 
Note that the expectation value is independent of the initial state $\rho'$, which we furthermore may restrict to the qubits that are in the support of $O_t$. We may further reduce the size of the circuit that needs to be implemented by restricting to the past causal cone. Combining these two observations leads to the statement of our main result:
\begin{theorem}\label{thm:effectivecircuit}
\normalfont
Let $O_T$ be an observable supported in a ball of radius $r$ and $\rho'$ be a state on the qubits that are in the support of $O_t$. It is possible to compute $\tr\lb \rho O_T\rb$ up to an additive error $2\delta(t,r)$ by implementing a circuit consisting of
$N_U(t,r)$ two-qubit gates on $N_Q(t,r)$ qubits.
\end{theorem}
\begin{figure}[h!]
\centering
\includegraphics[width=1\columnwidth,trim={0cm 0.2cm 7cm 5cm},clip]{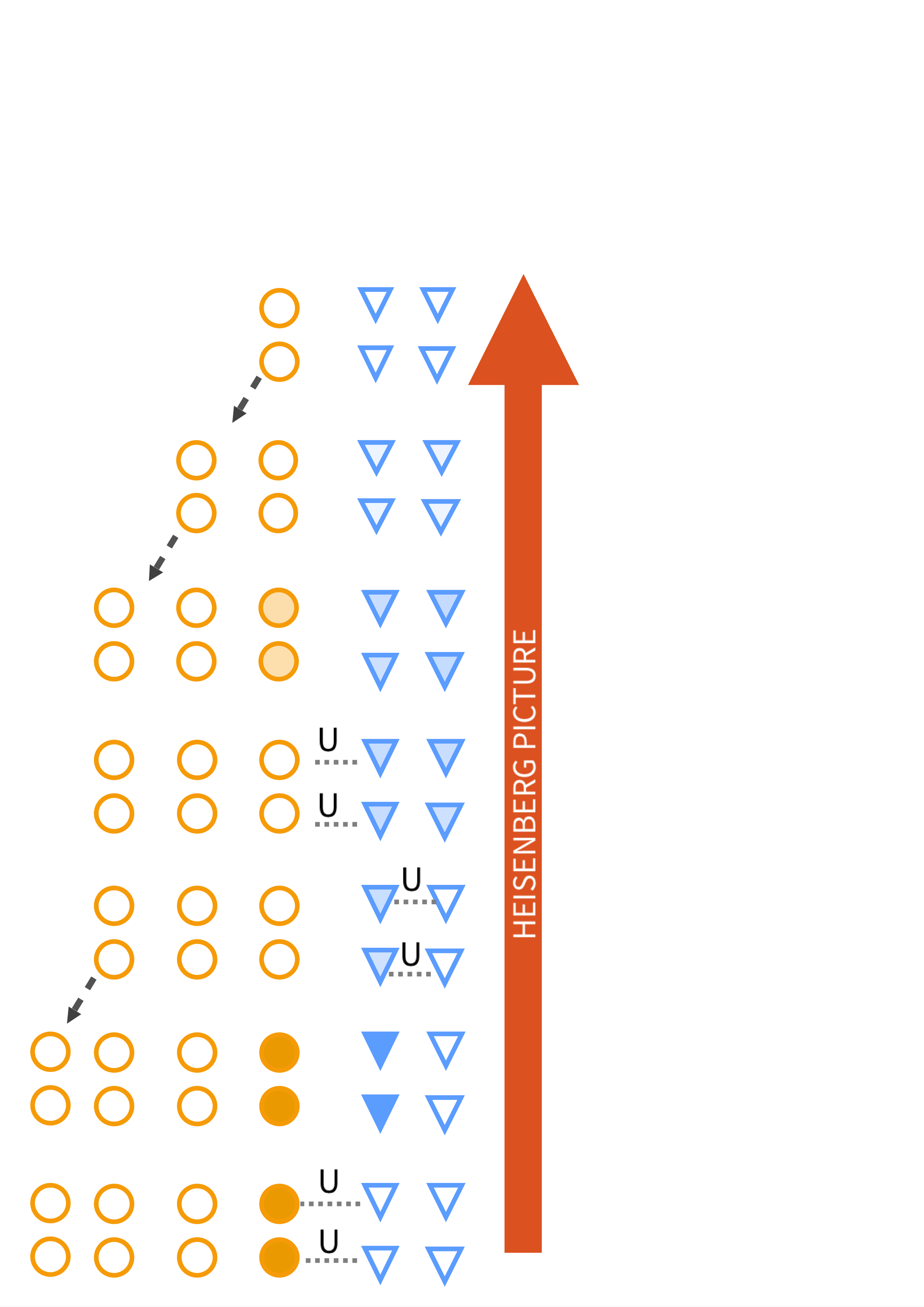}
\caption{(Mixing and support growth in the Heisenberg picture) : Evolution in the Heisenberg picture of an observable initially supported on the last two incoming qubits in the lower right corner (filled orange corner) at the fourth time step. The $U$ indicate that we apply a unitary between the involved qubits and the dashed arrows indicate that old qubits are shifted to the  left. Note that we suppressed the unitaries that do not  contribute to the expectation value. The distance of the observables to the identity is indicated by how filled the element, i.e. empty shapes indicate that the observable is proportional to the identity on that system. Note that as in the Heisenberg picture we discard qubits, this causes the observables to mix. Moreover, we see that after two iterations the observables are essentially proportional to the identity and it suffices to implement that part of the circuit to estimate it.\label{fig:2}}

\end{figure}

The theorem implies a way of performing VQE given bounds on $\delta(t,r)$ using a potentially smaller NISQ device than preparing the whole state. This is because implementing the smaller, effective circuit, requires fewer qubits and gates. This is illustrated in Fig.~\ref{fig:2}.

\subsection{Robustness to noise}
Consider the objective of calculating the ground state energy of a two local  Hamiltonian $H$, in the sense that it only acts on nearest neighbors in $G_T$, with each local term $H_i$ satisfying $\|H_i\|_\infty\leq1$. It suffices to estimate all $H_i$ individually to obtain an estimate of the global energy of the state by adding up the energy terms. Now suppose that we can implement each $2$-qubit gate with an error $\epsilon_U$ in operator norm and can prepare each initial qubit up to an error $\epsilon_P$. %
This implies that the total error of implementing the causal cone and measuring each $H_i$ is bounded by $\epsilon_U N_U(t,2)+\epsilon_P N_Q(t,2)$. Thus, by only implementing the circuit from iteration $t$ to $T$, it is possible to estimate the energy of each term with an error of
\begin{linenomath}
\begin{align}\label{equ:bounderror}
2\delta(t,2)+\epsilon_U N_U(t,2)+\epsilon_P N_Q(t,2).
\end{align}
\end{linenomath}
This generalizes to observables with arbitrary radius $r$ and can be improved to by exploiting the fact that :
\begin{linenomath}
\begin{align}
    &2\delta(t,r)+\sum_{k=t+1}^{T}\delta(k-1,r)\epsilon_U\lb N_U(k,r)-N_U(k-1,r)\rb\nonumber\\
    &+\sum_{k=t+1}^{T}\delta(k-1,r)\epsilon_P\lb N_Q(k,r)-N_Q(k-1,r)\rb.
\end{align}
\end{linenomath}
To see this, recall that $\delta(k,r)$ measures how close the operator $O_k$ is to being proportional to the identity, as there exists an operator $A_k$ with the same support as $O_k$ such that $O_k$ can be decomposed into $O_k=c\one+\delta(k,r)A_k$.
At the $k$'th iteration, any evolution in the Heisenberg picture only acts non-trivially on $A_k$ and changes the expectation value of the observable w.r.t.~to any state by at most $\delta(k,r)\|O_T\|$. Thus, if we actually implement a noisy version of the original evolution which is $\epsilon_U$ close to it, then we can only notice the effect of the noise in the part given by $\delta(k,r)A_k$. We conclude that each noisy unitary contributes with an error at most $\epsilon_U\delta(t,r)$, i.e. the effect of noise decreases in time if $\delta(t,r)$ decays. As there are $N_U(k,r)-N_U(k-1,r)$ new unitaries in the causal cone at the iteration, we obtain the bound. We note that the noise robustness  we obtain is of the same order as the one obtained implementing the whole circuit, as in Refs.~\cite{kimswingle,kim2017noise}, and refer to the Supplementary Information for a detailed discussion.

These results are related with the fact that $\delta(t,r)$ and the geometry of the interactions govern the correlations present in the state produced. For $E_T,F_T$ two observables of disjoint support of radius $r$ and $t$ be the largest $t$ such that $E_t$ and $F_t$ have supports that intersect we can show that:
\begin{linenomath}
\begin{align*}
    \left|\tr\lb \rho E_T\otimes F_T\rb-\tr\lb \rho E_T\rb\tr\lb \rho F_T\rb\right|\leq 6\delta(t,r).
\end{align*}
\end{linenomath}
As the decay of $\delta(t,r)$ also governs the noise robustness, we see that there is a trade-off between the correlation length and the robustness to noise. For instance, one should expect $\delta(t,r)$ to be exponentially decaying for states with finite correlation length.

\subsection{Estimating convergence}
It is necessary to bound $\delta(t,r)$ in our approach in order to bound the error. Thus, it is important to find conditions that guarantee the decay of the mixing rate and to develop protocols to estimate the mixing rate on a NISQ device. In the translationally invariant case, one can apply the large toolbox available to estimate mixing time bounds~\cite{Burgarth2013,Temme2010,Reeb2011,Bardet2017,Muller-Hermes2018}, as further explained in the Supplementary Information.

However, it is important to acknowledge that obtaining rigorous mixing time bounds is notoriously difficult even for classical systems~\cite{Levin_2017}. 
But this has not kept Markov Chain Monte Carlo algorithms from being one of the most successful methods to simulate physical systems~\cite{Binder_2019}. There exist many heuristic methods for classical systems~\cite{Cowles_1996} and here we also discuss a heuristic method to determine when the circuit has stabilized. As made transparent by Eq.~\eqref{equ:mixingandapprox1}, whenever the circuit has converged, the output of the circuit is independent of the initial state $\rho'$. Thus, one possible way of checking that the circuit has indeed converged is testing several different initial states and making sure that the expectation value of the output with respect to the observable does not depend on the input. This can be done by picking a set of initial states that is overcomplete, i.e. spans the space of all states as detailed in the Methods section. 

\begin{figure}[h!]
\centering

\includegraphics[width=0.75\columnwidth]{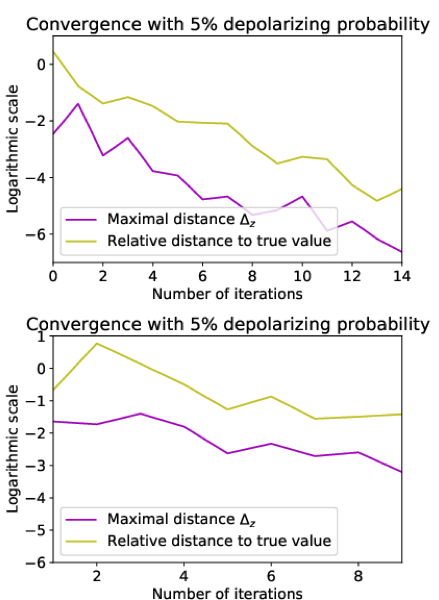}

\caption{(Convergence check): Illustration of convergence check for MPS in a rapid and slow mixing scenario. We consider a bath of $9$ qubits arranged on a line. In each iteration a new system qubit is added and a fixed circuit is run between the bath and the new system qubit. In the rapid mixing case, we perform a depth $3$ circuit between the system qubit and the bath such that each gate is followed by a depolarizing channel with probability 5\%. The complete state was evolved for $50$ time steps and, thus, consists of total of $50$ qubits, and we measured Pauli observables on the last two qubits. In the slow mixing we perform a depth $2$ circuit instead. If we denote by $X_{t}$ the expectation value of the state evolved by $t$ steps, we define the relative error to the true value to be $|1-X_{t}/X_{50}|$, as $X_{50}$ corresponds to preparing the whole state. To generate each plot, we have generated $50$ instances where the gates in the circuits were picked at random for each instance. The figures display the logarithm in base $10$ of the average of the quantities. Even with moderate levels of noise, we can faithfully reproduce the expectation value up to a relative error of $10^{-2}$ after $5$ iterations, giving an order of magnitude saving in the number of qubits in the rapid mixing scenario, while in the slow mixing, we required around $13$.
Moreover, we have picked $20$ random initial states to estimate $\Delta_{z}$, as more initial states did not seem to improve the estimate. As we can see, $\Delta_{z}$ is a good proxy for the relative distance to the true expectation value. \label{fig:convergence2}}

\end{figure}

In short, the approach is to pick random initial product states on the support of the observable $O_t$ and compare the value of the expectation value to that of the initial state where all qubits in the support are in state $\ket{0}$. If the expectation value with respect to different initial states all coincide, we build confidence that the computation has indeed converged. On the other hand, if the expectation values differ for two different initial states, then we have not converged and must go deeper and decrease $t$. We denote the maximal difference of the expectation value for the several different choices of initial state by $\Delta_z$ and refer to the methods section for a precise definition.
An example of the approach is shown in  Figure~\ref{fig:convergence2} for a matrix product states with both fast and slow convergence.

As we can see from the figures, estimating $\Delta_z$ gives reliable convergence diagnostics. Importantly, this is obtained with only a modest number of randomly selected input states. This suggests that estimating the convergence does not outweigh the over all advantage of bounding the circuit size for estimating local observables with our framework. We do note,  however, that this is a heuristic and not rigorous approach. For rigorously establishing convergence, it would in general require a sample complexity increasing exponentially with the support. We refer to the supplementary information for more details.

\begin{table} 
\centering
\begin{tabular}{| c| c | c | c |}
  \hline 
Scheme   & Error  & Gates & Qubits  \\
  \hline 
  DMERA  & $\epsilon_U \lambda^{-1}D^2+\epsilon_P \lambda^{-1}D$   & $t_\epsilon D^2$ & $t_\epsilon D$ \\
  \hline 
MPS & $\epsilon_U \lambda^{-2}D^2+\epsilon_P \lambda^{-1}D$ 
&  $t_{\epsilon}^2D^2 $ & $t_{\epsilon}D$\\
  \hline 
RI-$d$ &$\epsilon_U \lambda^{-d-1}D^{d+1}+\epsilon_P \lambda^{-d}D^d$ & $t_{\epsilon}^{d+1}D^{d+1}$ & $t_{\epsilon}D^{d}$\\
  \hline  
\end{tabular}
\caption{(Summary of resources): Error and resources required for implementing the effective causal cone, as in~(\ref{equ:mixingandapprox1}), for different interaction schemes under the assumption that $\delta(t,r)=ce^{-\lambda (T-t)}$. RI-$d$ refers to the case in which a $d-$dimensional bath interacts with a $d-$dimensional system at each iteration. All entries are only up to leading order in $D$, $T$ and $\lambda$. 
For $\lambda$ independent of system size, we see that it is possible to approximate all expectation values with quantum circuits of constant size.\label{tab:comparison}}

\end{table}

\section{Discussion}
To demonstrate the implications of our results, we summarize the noise-robustness and required number of gates and qubits in Table~\ref{tab:comparison} for some interaction schemes. 
We are able to significantly decrease the number of unitaries and qubits compared to the approach of  Refs.~\cite{kim2017noise,kimswingle}. This is because we only require the circuit corresponding to the past causal cone until it stabilizes to be implemented, in contrast to the the whole causal cone. 
Clearly, these results also imply that it is possible to approximate these expectation values classically if the resulting effective circuits are of a classically simulatable size. 

Our results provide an intuitive understanding of the stability of these computations. Each iteration contributes less to the value of expectation values, which implies that there is a small effective quantum circuit underlying the computation. Furthermore, the size of this circuit is related to the correlation length of the state and the effect of noise decreases proportionally to the correlations between regions.

Although rigorously testing at which iteration the circuit has converged might require exponential resources, we see that chosing a few random initial states and comparing the different expectation values provides useful guidance to check wether convergence has occurred. This shows that it is feasible to build confidence in the convergence and required depth with moderate resources using such heuristics. However, it would still be interesting to establish more rigorous protocols under suitable additional assumptions.

There are significant challenges in scaling up current qubit technologies~\cite{Monroe2013,Lekitsche2017, Kjaergaard2019}. The reduction in the number of qubits that we have shown above means that it may be possible to explore many-body quantum states with NISQ devices with substantially fewer qubits, potentially bringing such tasks into reach for current technology~\cite{Zhang2017,Bernien2017}. The possible reduction in the number of gates also reduces the necessary runtime of the circuits, which is important for hardware subject to qubit loss over time such as trapped atoms~\cite{Saffman2016}.   Note that the objective of this method is to expand the simulating capabilities of NISQ devices subject to strict hardware limitations. This is in contrast to other techniques, like measurement regrouping~\cite{McClean2016,verteletskyi2019measurement,jena2019pauli,huggins2019efficient,gokhale2019minimizing,crawford2019efficient}, that focus on optimizing resources given the ability to implement the whole circuit that prepares a given state.

\section{Methods}
\subsection{Proofs and checking mixing}
The main result of our work is based on the lemma in the main text. In order to prove this lemma, we have used a method based on viewing the generation as a quantum channel in the Heisenberg picture. The formal proof is given below.  
\begin{proof}
By the definition of $\delta(t,r)$, we see that
\begin{linenomath}
\begin{align}\label{equ:evolo}
 \Phi^{*}_{[t,T]}(O_T)=O_t=c\one+\delta(t,r)A,   
\end{align}
\end{linenomath}
where $A$ is some observable supported on $\operatorname{supp}(O_t)$ satisfying $\|A\|_\infty\leq \|O_T\|_\infty$.  As $\Phi^{*}_{[0,t-1]}$ is a quantum channel in the Heisenberg picture, $\Phi^*_{[0,t-1]}(\one)=\one$ and $\|\Phi^{*}_{[0,t-1]}\|_{\infty\to\infty}\leq 1$~\cite{contractwolf}. Thus,
\begin{linenomath}
\begin{align}\label{equ:furtherevol}
\Phi^*_{[0,t-1]}\circ\Phi^*_{[t,T]}(O_T)=c\one+\delta(t,r)\Phi^*_{[0,t-1]}\lb A\rb, 
\end{align}
\end{linenomath}
where $\|\Phi^*_{[0,t-1]}\lb A\rb\|_\infty\leq \|O_T\|_\infty$. We conclude that
\begin{linenomath}
\begin{align*}
&\left| \tr\lb \Phi_{[t,T]}\lb \rho'\rb O_T\rb-\tr\lb \rho O_T\rb\right|&\\
&=\left| \tr\lb \rho' \Phi^*_{[t,T]}\lb O_T\rb \rb-\tr\lb \rho_0\otimes \rho_B \Phi^*_{[0,T]}\lb O_T\rb\rb\right|&\\
&=\delta(t,r)\left| \tr\lb \rho' A \rb-\tr\lb \rho_0\otimes \rho_B \Phi^*_{[0,t-1]}\lb A\rb\rb\right| \\
&\leq 2\delta(t,r)\|O_T\|_\infty
\end{align*}
\end{linenomath}
by combining~\eqref{equ:evolo} and~\eqref{equ:furtherevol}. 
\end{proof}

Our results show that having an estimate for the mixing rate $\delta(t,r)$ allows to bound the number of qubits and cirucit size needed to estimate local observable on a many-body quantum state. While certain classes of states are known to have rapid mixing leading to fast convergence of the onservables (like ground states of gapped Hamiltonians), we have developed a heuristic method for estimating the convergence of local observables. The method relies on the observation that we can expand any density matrix using a basis of single qubit states. 
Let $\rho_{t}=\Phi_{[0,t]}(\rho_0\otimes\rho_b)$ be the state we obtain by evolving from time $0$ to $t$ and $O_T$ be defined as usual. 
Moreover, let $m=N_Q(r,t)$ be the number of qubits in the support of $O_t$. To check the convergence of the circuit, we prepare input states
$\ket{\psi_{z}}=\bigotimes_{i=1}^m \ket{z_i}$, where $\ket{z_i}\in\{\ket{0},\ket{1},\ket{+},\ket{-}\}$. The states $\ket{\psi_{x}}$ thus corresponds to various product state combinations of the states $\ket{0},\ket{1},\ket{+}$ and $\ket{-}$ on the support of the observable $O_t$. It is well-known that these states form a basis of the space of Hermitian matrices. 
We can therefore expand $\rho_t$ as 
\begin{linenomath}
\begin{align*}
\rho_t=\sum\limits_{z}a_{z}\ketbra{\psi_{z}}{\psi_{z}},
\end{align*}
\end{linenomath}
where $a_{z}\in\mathbb{C}$. Furthermore, it is easy to see that $\sum_{z}a_{z}=1$ by taking the trace. Now define $\Delta_{z}$ to be  given by:
\begin{linenomath}
\begin{align*}
\Delta_{z}=\tr((\ketbra{\psi_{z}}{\psi_{z}}-\ketbra{0}{0}^{\otimes m})O_t).
\end{align*}
\end{linenomath}
From this, we immediately obtain that:
\begin{linenomath}
\begin{align}\label{equ:expansion}
\tr(\rho_tO_t)=\tr\lb\ketbra{0}{0}^{\otimes m}O_t\rb+\sum_{z}a_{z}\Delta_{z}.
\end{align}
\end{linenomath}
Eq.~\eqref{equ:expansion} suggests the simple protocol of picking random initial product states $\ket{\psi_{z}}$ and comparing the expectation value with the outcome for initial state $\ket{0}^{\otimes m}$ to estimate the convergence. If the expectation value with respect to different initial states$\ket{\psi_{z}}$ all coincide with the one of  $\ket{0}^{\otimes m}$, we build confidence that all $\Delta_{z}$ are small, thus ensuring that the expectation value is similar as picking the initial state to be $\ket{0}^{\otimes m}$.

\section{Acknowledgements}
We would like to thank Albert H. Werner, Carlos E. Gonz\' alez-Guill\' en and Giacomo de Palma for helpful discussions. We acknowledge financial support from the VILLUM FONDEN via the QMATH Centre of Excellence (Grant no. 10059),  from the European Research Council (ERC Grant Agreement No. 337603) and
from the QuantERA ERA-NET Cofund in Quantum Technologies implemented within the European Union's Horizon 2020 Programme (QuantAlgo project) via the Innovation Fund Denmark. 
JB acknowledges funding from the NWO Gravitation Program Quantum Software Consortium.

\bibliographystyle{naturemag}
\bibliography{biblioa}

\setcounter{equation}{0}
\setcounter{figure}{0}
\setcounter{table}{0}
\makeatletter
\renewcommand{\theequation}{S\arabic{equation}}
\renewcommand{\thefigure}{S\arabic{figure}}
\renewcommand{\bibnumfmt}[1]{[S#1]}
\renewcommand{\citenumfont}[1]{S#1}
\end{document}


\begin{center}
\textbf{\large Supplementary Material}
\end{center}
This is the supplementary material to the article \emph{Noise-robust exploration of quantum matter on near-term quantum devices}. We first discuss in more detail the growth of the causal cone, the number of unitaries, and error estimates for the examples considered in the article  (Sec.~\ref{sec:causalcone}). We then review the connection between mixing times of quantum channels and the decay of the mixing rate function (Sec.~\ref{sec:mixingrates}). Here, we also show that the mixing rate and the geometry of the interaction scheme bound the correlation length of sequentially generated states. Finally, we elaborate on the comparison to the results of Refs.~\cite{kim2017noise,kimswingle} (Sec.~\ref{app:resultskim}) and describe a protocol to certify that a circuit is mixing for a given observable (Sec.~\ref{sec:mixing}).

\section{Supplementary Discussion}
\subsection{Causal cone of DMERA and sequentially generated states}\label{sec:causalcone}
In this subsection, we review the constructions for the examples considered in the main article, analyse the growth of the past causal cone and the corresponding implications for the scaling of the error of noisy implementations.

Let us start by briefly recalling for the reader's convenience the construction of DMERA states given in Ref.~\cite{kimswingle}, which are depicted in Supplementary Figure~\ref{fig:interactionschemedmera}, .
We start with a system consisting of one qubit. Then, at iteration $t$ we add $2^{t-1}$ new qubits to the system, placing one qubit to the right of each existing qubit. 
Furthermore, at each iteration, we apply $D$ layers of two-qubit unitary gates between neighboring qubits. The resulting state has a final number of $2^T$ qubits and it is necessary to implement $(D-1)\lb2^{T+1}-1\rb$ two-qubit gates to prepare the whole state.

While we add $2^{t-1}$ qubits in the Schr\"oedinger picture, when looking at the Heisenberg picture of the evolution we will discard half of the qubits at each iteration. This ensures that the dynamics in the Heisenberg picture will typically be locally mixing. However, as it is the case for usual MERA, local observables have by design a causal cone that is of polynomial size in $t$, which is crucial to all estimates in the main article. We will now discuss their growth in more detail.

\begin{figure}[h!]
\includegraphics[scale=0.5,trim={0cm 16cm 0cm 1cm},clip]{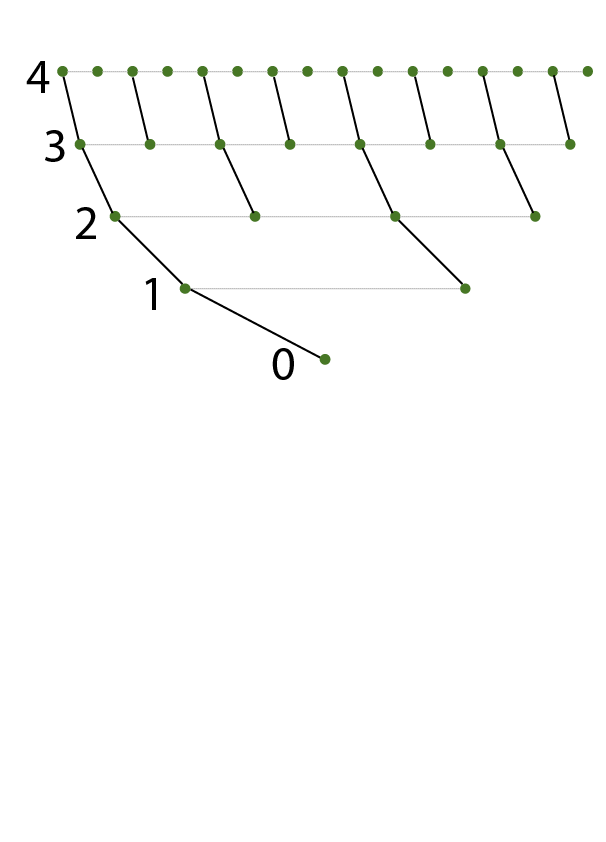}
\caption{Depiction of the DMERA for iterations $0$ to $4$. The circles (green, filled) denote the system qubits. The thick, black lines indicate where a qubit goes from one iteration to the next and the thin, gray lines indicate which qubits are neighbors at a given iteration. The digits, always next to the first qubit, indicate the iteration. }
\label{fig:interactionschemedmera}
\end{figure}

Let us start with the number of unitaries in the past causal cone in DMERA. Recall that when looking at what happens at each iteration in the Heisenberg picture, after discarding every second qubit present in the previous iteration, we apply a unitary circuit of $D$ layers, always with the restriction that we can only apply unitaries between qubits that are neighbors on the line. When we apply the first layer, only unitaries which act on at least one qubit in the support have a nontrivial effect. Let $R(O_t)$ be the radius of the observable before we apply the first layer of unitaries. Then there at most $2R(O_t)-1$ nontrivial unitaries acting on the qubits in the support and two unitaries, one to the left of the support and one to the right, that act on the qubit in the left corner of the support and the first qubit to the left of the support and analogously to the right. Thus, we conclude that as we apply the first layer, we have $2R(O_t)+1$ unitaries acting nontrivially and the support will increase to one qubit to the right and one qubit to the left.
The next layer of the unitary circuit will then act on an observable of support with radius at most $R(O_t)+1$. Applying the same reasoning as before, we see that the total number of unitaries that act nontrivially is $2R(O_t)+3$.
We conclude that the total number of unitaries that acts nontrivially after repeating this process $D$ times is bounded by:
\begin{align}\label{equ:numberofunitariesintermsradius}
    \sum_{k=0}^{D-1}(2R(O_t)+2k+1)=D(2R(O_t)+D).
\end{align}
Let us now estimate the size of the radius at each iteration to obtain a more concrete bound on the number of unitaries. 

As we observed above, if at the beginning of an iteration the radius is $R(O_t)$, it will increase by $D$ and then be halved after we discard the qubits. Thus, it will go from $R(O_t)$ to at most $\ceil{(R(O_t)+D)/2}\leq(R(O_t)+D)/2+1$. Applying this recursive relation, we see that if the initial radius is $R(O_T)$ then at iteration $t$, the radius is bounded by 
\begin{align*}
   R(O_t)\leq R(O_T)2^{-(T-t)}+\sum\limits_{k=t}^T\frac{D+2}{2^{T-k}}=R(O_T)2^{-(T-t)}+(D+2)(2-2^{-(T-t)}). 
\end{align*}
Note that this implies that the radius of an observable is bounded by a constant independent of $t$.
Combining the bound above on the radius of the observable with Eq.~\eqref{equ:numberofunitariesintermsradius}, we obtain that the number of unitaries added to the cone at iteration $t$ is bounded by:
\begin{align}\label{equ:numberunititerationmera}
D(2R(O_T)2^{-(T-t)}+2(D+2)(2-2^{-(T-t)})+D).
\end{align}
From this we can easily bound  the total number of unitaries in the past causal cone from iteration $t$ to $T$ by summing the contribution at each step:
\begin{align} \label{eq:radius}
&N_U(t,R(O_T))\leq \sum\limits_{k=t}^TD(2R(O_k)+D)\leq \sum\limits_{k=t}^T D(2R(O_T)2^{-(T-k)}+2(D+2)(2-2^{-(T-k)})+D)\\&\leq (T-t)D(2R(O_T)+5D+8).\nonumber
\end{align} 
Let us now estimate the number of qubits in the past causal cone.
At every iteration, we grow the support by at most $D$ new qubits to the left and $D$ to the right, and we start with at most  $2R(O_T)$ qubits. This leads to the bound
\begin{align}\label{equ:numberofqubitsdmera}
N_Q(t,R(O_T))\leq 2R(O_T)+2D(T-t)
\end{align}

We will now estimate the error of implementing the past causal cone from iteration $t$ to $T$, which, as explained in the main text, can be bounded by:
\begin{align}\label{equ:ourstabilitybound}
    2\delta(t,r)+\sum_{k=t+1}^{T}\delta(k,r)\left[\epsilon_U\lb N_U(k,r)-N_U(k+1,r)\rb+\epsilon_P\lb N_Q(k,r)-N_Q(k+1,r)\rb\right],
\end{align}
where we assume that each unitary is implemented with an error of $\epsilon_U$ in the $1\to1$ norm~\footnote{strictly speaking, a bound in the diamond norm is required. However, as we will discuss in more detail later, as we only consider two qubit unitaries, they are related by a factor of four.} and we can initialize each qubit up to an error $\epsilon_P$, in the sense that we can prepare a state that is $\epsilon_P$ close in trace distance to the ideal one.  

Let us start by estimating the error stemming from the noisy unitaries. Note that the term $ N_U(k,r)-N_U(k+1,r)$ is nothing but the newly added unitaries at iteration $k$, which we bounded in Eq.~(\ref{equ:numberunititerationmera}).
It follows that the contribution to the error from the noisy unitaries from iteration $t$ to $T$ is bounded by
\begin{align}\label{equ:boundfrommaintext}
    \epsilon_U\sum\limits_{k=t}^T\left[N_U(k,R(O_T))-N_U(k+1,R(O_T))\right]\delta(k,R(O_T))\leq\epsilon_U\sum\limits_{k=t}^T D(2R(O_T)+5D+8)\delta(k,R(O_T)).
\end{align}
To illustrate the bound, we assume that $\delta(k,r)=e^{-\lambda (T-k)}$. Consequently, 
\begin{align}\label{equ:errorunitariesmera}
    \epsilon_U \sum\limits_{k=t}^T D(2R(O_T)+5D+8)e^{-\lambda (T-k)}= \epsilon_U \frac{D(e^\lambda-e^{-(T-t)\lambda})(2R(O_T)+5D+8)}{e^{\lambda}-1}.
\end{align}

One can do a similar computation for state preparation errors. As discussed above, at most $2D$ qubits are added to the causal cone for each iteration. Thus, the error caused by initialization between iterations $t$ and $T$ is bounded by:
\begin{align}\label{equ:errorsprepmera}
\epsilon_P\lb2R(O_T)+2D\sum_{k=t}^T e^{-\lambda(T-t)}\rb=\epsilon_P\lb 2R(O_T)+2D\frac{ \left(e^{\lambda}-e^{-(T-t)\lambda}\right)}{e^\lambda-1}\rb.
\end{align}
From combining equations~\eqref{equ:errorunitariesmera} and~\eqref{equ:errorsprepmera}, we can conclude that the error in estimating the expectation value of an observable  by implementing the past causal cone from iteration $t$ to $T$ is bounded by:
\begin{align*}
    &\epsilon_U \frac{D(e^\lambda-e^{-(T-t)\lambda})(2R(O_T)+5D+8)}{e^{\lambda}-1}+
   \epsilon_P\lb 2R(O_T)+2D\frac{ \left(e^{\lambda}-e^{-(T-t)\lambda}\right)}{e^\lambda-1}\rb+2e^{-\lambda(T-t)}.
\end{align*}
Let us now suppose we only implement the past causal cone from iterations $t_{\epsilon_U}=T-\lambda^{-1}\log(\epsilon_U^{-1})$ until $T$.
The resulting error will then be at most
\begin{align*}
    &\epsilon_U \lb\frac{D(e^\lambda-\epsilon_U)(2R(O_T)+5D+8)}{e^{\lambda}-1}+2\rb+
   \epsilon_P\lb 2R(O_T)+2D\frac{ \left(e^{\lambda}-\epsilon_U\right)}{e^\lambda-1}\rb.
\end{align*}
By approximating $(e^{\lambda}-1)^{-1}$ by $\lambda^{-1}$, we see that the error stemming from the noisy unitaries is at most of order $\epsilon_U D^2\lambda^{-1}$. Similarly, the error from noisy initialization of qubits is at most of order $\epsilon_P\lambda^{-1}D$.
Moreover, by inserting $t_{\epsilon_U}$ into Eq.~\eqref{equ:numberofqubitsdmera}, we obtain that the total number of qubits necessary to perform this computation is at most
\begin{align*}
    2R(O_T)+2D\lambda^{-1}\log(\epsilon_U^{-1})
\end{align*}
and the number of unitaries that needs to be implemented is bounded by
\begin{align*}
    \lambda^{-1}\log(\epsilon_U^{-1})D(2R(O_T)+5D+8),
\end{align*}
which follows from inserting $t_{\epsilon_U}$ into Eq.~\eqref{eq:radius}. Thus, under these assumptions it possible to compute local expectation values of fixed radius with noisy circuits whose error and size only depends on $\lambda$ and $D$, not $T$.

Another important subclass of states are those that are sequentially generated.
The most prominent example is matrix product states (MPS). Here, only one qubit interacts with the bath at each iteration. 
A simple generalization of this is where a group of qubits (arranged according to a $d-$dimensional graph) interacts with a bath (arranged according to a $d'$-dimensional graph) at each iteration, see Supplementary Figure~\ref{fig:repeatedinteraction} for an example with $d=0$ and $d'=2$. For this work, we also make the restriction that the interaction is given by a circuit of depth at most $D$. Setting $d=0$ and $d'=1$, i.e. a qubit interacting with qubits on a line, recovers our version of MPS. We will also discuss the case of $d=d'$ in more detail, which we will refer to as RI-$d$. The of case $d=d'=1$ encapsulates examples like holographic computation discussed in~\cite{kim2017b}. 
\begin{figure}[!tbp]
  \centering
  \begin{minipage}[b]{0.3\textwidth}
    \includegraphics[width=\textwidth,trim={0cm 9cm 0cm 1cm},clip]{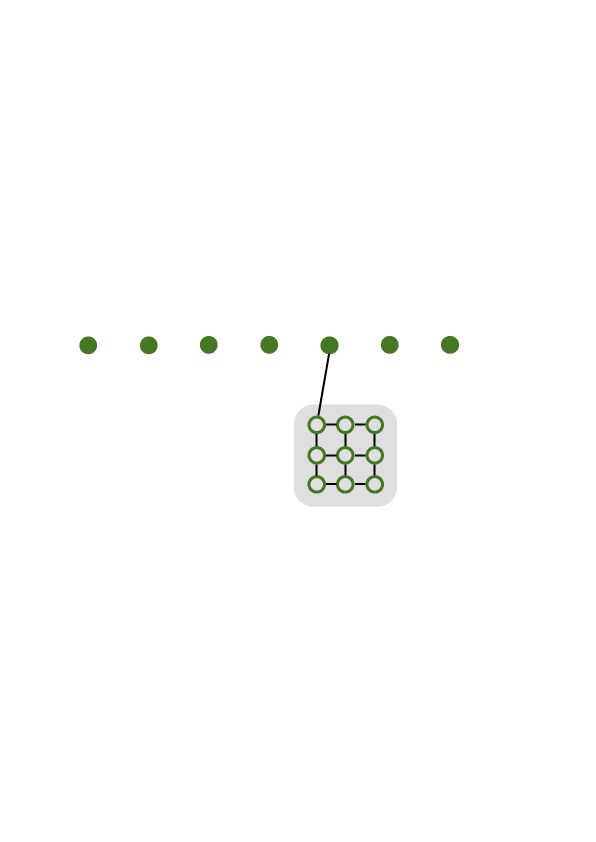}
  
  \end{minipage}
  \begin{minipage}[b]{0.3\textwidth}
    \includegraphics[width=\textwidth,trim={0cm 9cm 0cm 1cm},clip]{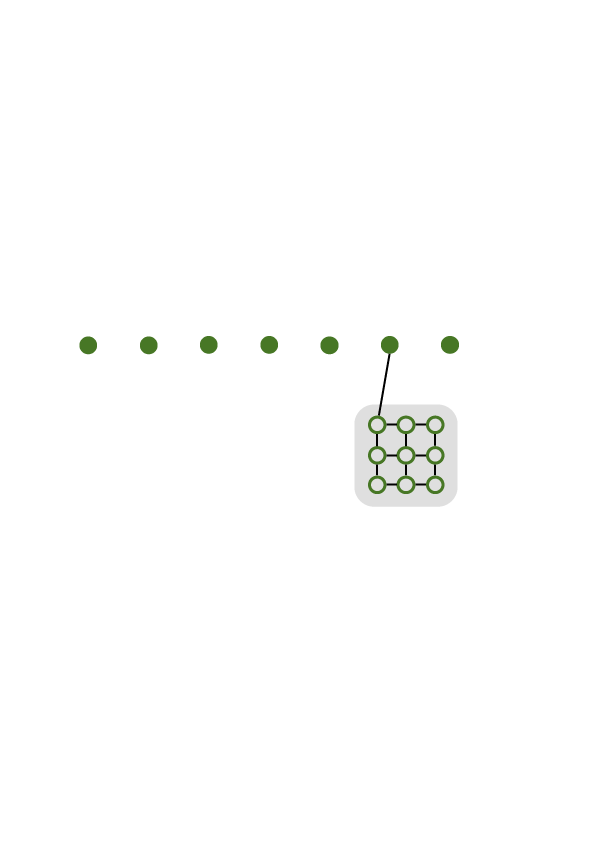}
   
  \end{minipage}
  \begin{minipage}[b]{0.3\textwidth}
    \includegraphics[width=\textwidth,trim={0cm 8cm 0cm 12cm},clip]{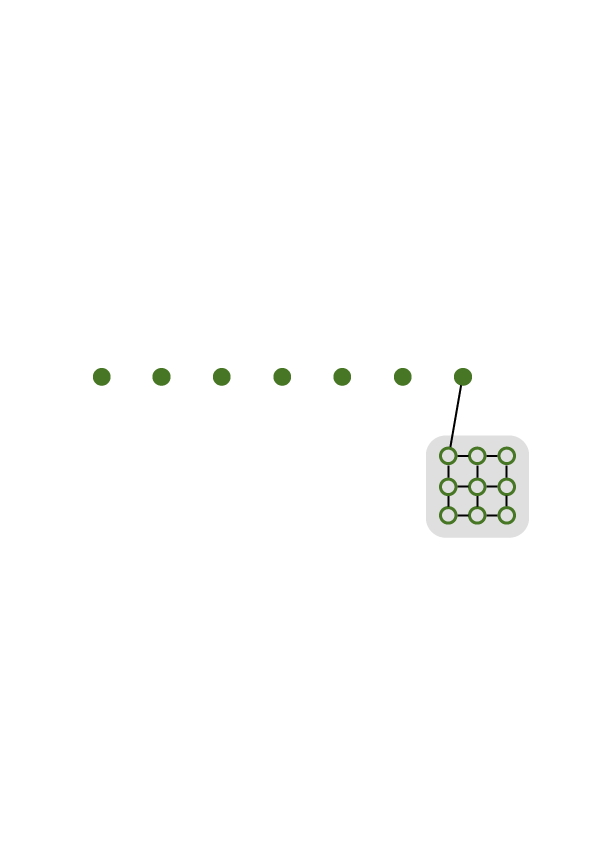}
    
  \end{minipage}
  \caption{Three subsequent iterations of a repeated interaction system with $d=0$ and $d'=2$. That is, the system qubits (green, filled circles) interact at each iteration with bath qubits (green, empty circles) that are arranged according to a two dimensional graph. The black, thick line indicates which system qubits is interacting with the bath, while the bath qubits interact with nearest neighbors.}
  \label{fig:repeatedinteraction}
\end{figure}

We now discuss the growth of scaling of the errors in both our version of MPS and RI-$d$. Unlike we did for DMERA, we will not fix the exact graph that models the interactions in the bath and between system and bath at each iteration and choose to focus on the scaling of the size of causal cones. More precisely, we will assume that there are constants $C_V$ and $C_E$ such that for every ball of radius $r$ in the interaction graph there are at most $C_Er^d$ edges and $C_Vr^d$ vertices inside the ball.

Let us now analyse the growth of causal cones. As it was the case with DMERA, if at the beginning of iteration $t$ the radius of an observable is $R(O_t)$, it will then grow to at most $R(O_t)+D$. However, unlike for DMERA, for the interaction schemes considered here we do not discard qubits between different iterations. Thus, the radius at iteration $t$ of an observable is bounded by $R(O_T)+(T-t)D$. This allows us to conclude that the number of qubits in the past causal cone is bounded by:
\begin{align*}
    N_Q(t,R(O_T))\leq C_V(R(O_T)+(T-t)D)^d
\end{align*}
for RI-$d$ and $C_V(R(O_T)+(T-t)D)$ for MPS.
Let us now do a similar computation for the number of unitaries in the past causal cone.
Supposing that the radius of the observable is $R(O_t)$ at the beginning of the iteration, there are at most $C_E(R(O_t)+1)^d$ unitaries that act nontrivially on the first layer and the radius will grow by one. For the second layer, there will be at most $C_E(R(O_t)+2)^d$ and the radius will again grow by one. We conclude that applying the $D$ layers will require a total of at most
\begin{align}\label{equ:numberunitaries}
    C_E\sum\limits_{k=0}^{D-1}(R(O_t)+k+1)^{d}
\end{align}
unitaries for iteration $t$. As $(R(O_t)+k+1)^{d}$ is monotone increasing in $k$, we have that the number of unitaries added at each iteration is bounded by:
\begin{align*}
    &C_E\sum\limits_{k=0}^{D-1}(R(O_t)+k+1)^{d}\leq C_E\int\limits_{1}^{D}(R(O_t)+x+1)^{d}dx
    = \frac{C_E}{d+1}\left[\lb R(O_t)+D+1\rb^{d+1}-\lb R(O_t)+2\rb^{d+1}\right]\\
    \leq &\frac{C_E}{d+1}\left[\lb R(O_T)+(T-t+1)D+1\rb^{d+1}-\lb R(O_T)+(T-t)D+1\rb^{d+1}\right],
\end{align*}
where for the last inequality we used our estimate for the radius of the observable at iteration $t$ and the fact that the function $x\mapsto \lb x+D+1\rb^{d+1}-\lb x+2\rb^{d+1}$ is monotone increasing for $x\geq0$ and $D\geq1$, as can be seen by a direct inspection of its derivative.
Thus, we can bound the maximum number of unitaries in the causal cone between iteration $t$ and $T$ by:
\begin{align}\label{equ:numberofunitariesinfinalRI}
    N_U(t,r)\leq \frac{C_E}{d+1}\sum_{k=t}^T\left[\lb R(O_T)+(T-k+1)D+1\rb^{d+1}-\lb R(O_T)+(T-k)D+1\rb^{d+1}\right].
\end{align}
Let us now estimate this sum. To this end, define the function $f(x)=\lb R(O_T)+xD+1\rb^{d+1}$. By the mean value theorem there exists $\xi_k\in[T-k,T-k+1]$ such that:
\begin{align}\label{equ:newunitariesRIdperstep}
    f(T-k+1)-f(T-k)=f'(\xi_k)=(d+1)D\lb R(O_T)+\xi_kD+1\rb^{d}\leq(d+1)D\lb R(O_T)+(T-k+1)D+1\rb^{d}.
\end{align}
Thus, inserting this bound into~\eqref{equ:numberofunitariesinfinalRI} it follows that  
\begin{align*}
    &N_U(t,r)\leq C_E(d+1)\sum_{k=t}^TD\lb R(O_T)+(T-k+1)D+1\rb^{d}\leq (d+1)C_ED\int\limits_{1}^{(T-t)+2}\lb R(O_T)+xD+1\rb^ddx\\
    &=C_E\left[\lb R(O_T)+(T-t+2)D+1\rb^{d+1}-\lb R(O_T)+D+1\rb^{d+1}\right].
\end{align*}
In particular, for MPS this gives a bound of 
\begin{align*}
    N_U(t,r)\leq C_E\left[\lb R(O_T)+(T-t+2)D+1\rb^{2}-\lb R(O_T)+D+1\rb^{2}\right].
\end{align*}

We now assume that $\delta(t,r)=e^{-(T-t)\lambda}$ to bound the estimation error from implementing the past causal cone, as we did with DMERA.
Recall that we bounded the number of new unitaries in the past causal cone at each iteration in Eq.~\eqref{equ:newunitariesRIdperstep}. Once again, combining these estimates with our assumption on the mixing rate function and~\eqref{equ:ourstabilitybound} yields a bound on the error stemming from the unitaries of at most
\begin{align*}
   \epsilon_U C_E\sum\limits_{k=t}^T D(d+1)\lb R(O_T)+(T-k+1)D+1\rb^{d}e^{-\lambda(T-k)}.
\end{align*}
Let us now estimate this sum. First, define the function
\begin{align*}
    g(x)=\lb R(O_T)+xD+1\rb^{d}e^{-\lambda x}.
\end{align*}
We have:
\begin{align*}
    g'(x)=\lb R(O_T)+xD+1\rb^{d-1}e^{-\lambda x}\lb dD-\lambda\lb R(O_T)+xD+1\rb\rb
\end{align*}
For $x\geq0$, we see that the function is monotone increasing for 
\begin{align*}
    x\leq x_0:=\frac{1}{D}\lb \frac{dD}{\lambda}-R(O_T)-1\rb
\end{align*}
and monotone decreasing for $x\geq x_0$. This allow us to conclude that:
\begin{align}\label{equ:inequalityintegral}
    &\epsilon_U C_E\sum\limits_{k=t}^T D(d+1)\lb R(O_T)+(T-k+1)D+1\rb^{d}e^{-\lambda(T-k)}\leq \epsilon_UC_ED(d+1)\left[\int\limits_{0}^{\ceil{x_0}}g(x)dx+\int\limits_{\lfloor x_0\rfloor}^{T-t+1}g(x)dx\right]\\
    &\leq2\epsilon_UC_ED(d+1)\int\limits_{0}^{T-t+1}g(x)dx.
\end{align}
It now remains to estimate this integral.
It is easy to compute the integral above using integration by parts $d$ times, although the resulting expressions are quite involved. We only reproduce them for $d=1$ and $d=2$ here.
For $d=1$ we have:
\begin{align}\label{equ:estimatefordone}
     \int\limits_{0}^{T-t+1}(R(O_T)+(x+1)D)e^{-\lambda x}dx=\frac{1-e^{-\lambda(T-t+1)}}{\lambda^2}\lb \lambda R(O_T)+D\lambda+D\rb-\lambda^{-2}e^{-\lambda(T-t+1)}\lb D\lambda(T-t+1)\rb,
\end{align}
and for $d=2$ we obtain:
\begin{align}\label{equ:estimatefordtwo}
    &\frac{1-e^{-\lambda(T-t+1)}}{\lambda^3}\lb \lambda^2(D+R(O_T))^2+2\lambda D(D+R(O_T))+2D^2\rb\\
    &-\lambda^{-3}e^{-\lambda(T-t+1)}\lb 2\lambda^2D(R(O_T)+D)(T-t+1)+D^2\lambda^2(T-t+1)^2\rb.\nonumber
\end{align}
It is then possible to obtain explicit bounds by combining the equations above with Eq.~\eqref{equ:inequalityintegral}. But it is easy to see by direct inspection that, assuming $R(O_T)\leq D$, the error will converge exponentially fast in $(T-t)$ to $\mathcal{O}\lb\epsilon_UC_E \lb D\lambda^{-1}\rb^{d+1}\rb$, which is again independent of $T$.  
It is also possible to obtain more explicit bounds on the asymptotic behaviour of the error, i.e. with $T\to\infty$. 
To this end, note that $g(x)\geq0$, thus:
\begin{align*}
    &\int_{0}^{T-t+1}g(x)dx\leq\int_{0}^{+\infty}\lb R(O_T)+xD+1\rb^{d}e^{-\lambda x}dx=D^{-1}\int\limits_{R(O_T)+1}^{\infty}y^de^{-\frac{\lambda}{D}\lb y-R(O_T-1)\rb}dy\\
    &\leq D^{-1}\int\limits_{0}^{\infty}y^de^{-\frac{\lambda}{D}\lb y-R(O_t-1)\rb}dy=\frac{e^{\frac{\lambda}{D}\lb R(O_T)+1\rb}}{D}\int\limits_{0}^{+\infty}y^de^{-\frac{\lambda}{D}\lb y\rb}dy=\frac{D^de^{\frac{\lambda}{D}\lb R(O_T)+1\rb}}{\lambda^{d+1}}\int\limits_{0}^{+\infty}z^{d}e^{-z}dz=\frac{d!D^de^{\frac{\lambda}{D}\lb R(O_T)+1\rb}}{\lambda^{d+1}}.
\end{align*}
This allows us to conclude that the noise stemming from the noisy unitaries is bounded by:
\begin{align*}
   2C_E\epsilon_U \frac{(d+1)!D^{d+1}e^{\frac{\lambda}{D}\lb R(O_T)+1\rb}}{\lambda^{d+1}}.
\end{align*}
Similar estimates hold for the total initialization errors ($\epsilon_P$).
We see that the number of qubits added at iteration $t$ is bounded by:
\begin{align*}
    (R(O_T)+(T-t+1)D)^d-(R(O_T)+(T-t)D)^d\leq dD(R(O_T)+(T-t+1)D)^{d-1},
\end{align*}
again using the mean value theorem. Thus, we may estimate the initialization error by:
\begin{align}\label{equ:estimatequbiterror}
    \epsilon_PC_VdD\sum\limits_{k=t}^{T}(R(O_T)+(T-k+1)D)^{d-1}e^{-\lambda t}.
\end{align}
The attentive reader must have already realized that the expression in~\eqref{equ:inequalityintegral} coincides with that of~\eqref{equ:estimatequbiterror} up to a constant if replace $d+1$ by $d$. Thus, we may use the same estimation techniques and conclude that the error is bounded by $\mathcal{O}\lb\epsilon_PC_V \lb D\lambda^{-1}\rb^{d}\rb$. Moreover, we may resort to the expressions in~\eqref{equ:estimatefordone} and~\eqref{equ:estimatefordtwo} if more refined inequalities in terms of $t$ and $R(O_T)$ are desired.
Thus, the total error of implementing the causal cone from $T-t$ to $t$ is bounded by:
\begin{align*}
    2C_V\epsilon_P \frac{d!D^{d}e^{\frac{\lambda}{D}\lb R(O_T)+1\rb}}{\lambda^{d}}+2C_E\epsilon_U \frac{(d+1)!D^{d+1}e^{\frac{\lambda}{D}\lb R(O_T)+1\rb}}{\lambda^{d+1}}+2e^{-\lambda(T-t)},
\end{align*}
up to corrections that are exponentially small in $T-t$.

\subsection{Mixing rates of quantum channels}\label{sec:mixingrates}
\newcommand{\M}{\mathcal{M}}
In this subsection, we clarify the connections between the mixing rate function and the mixing properties of quantum channels~\cite{Burgarth2013}. 
\begin{definition}[Mixing quantum channel]
A quantum channel $\Lambda:\M_d\to\M_d$ is called \textbf{mixing} if there is a unique state $\sigma$ such that $\Lambda(\sigma)=\sigma$ and for all states $\rho$ we have that
\begin{align*}
\lim\limits_{n\to\infty}\Lambda^n(\rho)=\sigma,
\end{align*}
where $\Lambda^n$ denotes the quantum channel composed with itself $n$ times.
\end{definition}
Given a mixing quantum channel $\Lambda$, the main quantity of interest is $t_1(\epsilon)$, defined as
\begin{align*}
t_1(\epsilon)=\inf\{n|\sup_{\rho}\|\Lambda^{n}\lb \rho\rb-\sigma\|_1\leq \epsilon\}.
\end{align*}
For $\epsilon>0$ this quantity measures how long it takes for the quantum channel to converge, i.e., its mixing time~\cite{Burgarth2013,Temme2010}.
Here $\|\cdot\|_1$ corresponds to the trace norm.
It is well-known that correlations in tensor network or finitely correlated states are governed by mixing properties of the transfer operator~\cite{Fannes1992,Perez-Garcia2006}. We will now show this connection for completeness of the exposition.

Note that
\begin{align*}
\sup\limits_{\rho}\|\Lambda^n(\rho)-\sigma\|_1
\end{align*}
corresponds to the $1\to1$ norm of the linear operator $\Lambda-\Lambda_{\infty}$, where $\Lambda_\infty(\rho)=\tr(\rho)\sigma$. It follows from duality that:
\begin{align*}
\|\Lambda^n-\Lambda_{\infty}\|_{1\to 1}=\|(\Lambda^n)^*-\Lambda_{\infty}^*\|_{\infty\to \infty}
\end{align*}
and $\Lambda^*_{\infty}(O)=\tr(\sigma O)\one$. 
Now suppose, for simplicity, that we wish to compute the expectation value of an observable $O$ supported on one qubit in $S_T$ and our interaction scheme is that of MPS. In this case, the qubits only interact with the bath at each iteration and not each other. Moreover, let us assume that the system is translationally invariant in the sense that we assume that $\mathcal{U}_t$ is the same for all $t$. Now note that 
\begin{align*}
O_T=\Phi_{T}^*(O)=\tr_{S_T A_T}\lb \mathcal{U}_{T}^*\lb O\otimes \one_{S_1\ldots S_{T-1}S_B}\rb\rb.
\end{align*}
will be an observable supported on the bath alone. Furthermore,
\begin{align*}
\Phi_{t}^*(O_{t+1})=\tr_{S_{t}A_{t}}\lb \mathcal{U}_{t}^*\lb O_{t+1}\otimes \one_{S_1\ldots S_{t}}\rb\rb.
\end{align*}
Since we have assumed the action of all $\mathcal{U}_{t}$ to be the same, we may define the quantum channel $\Lambda_{B}^*$ from the bath to itself as
\begin{align*}
\Lambda_{B}^{*}(X)=\tr_{S_{t}A_{t}}\lb \mathcal{U}_{t}^*\lb X\otimes \one_{S_1\ldots S_{t}}\rb\rb.
\end{align*}
We then have that $O_t=\lb \Lambda_B^*\rb^{T-t}(O_1)$. 
If $\Lambda_B$ is mixing, which is the generic case~\cite{Burgarth2013}, we may directly bound the mixing rate with a mixing time bound on $\Lambda_B$. Let 
\begin{align*}
\mathcal{B}_r=\{O:R(O)\leq r,\|O\|_\infty\leq1\}.
\end{align*}
Observe that
\begin{align*}
\delta(t,r)=\sup\limits_{O\in \mathcal{B}_r}\inf_{c\in \R}\|\Phi^*_{[t,T]}\lb O\rb-c\one\|_\infty=
\sup\limits_{O\in\mathcal{B}_r}\inf_{c\in \R}\| \lb \Lambda_B^*\rb^{T-t}(O)-c\one\|_\infty.
\end{align*}
For $\Lambda_B$ mixing, a natural choice for the constant $c$ is given by $\tr\lb O_1 \sigma\rb$, as in this case we have:
\begin{align*}
\delta(t,r)\leq \sup\limits_{\mathcal{B}_r}\| \lb \Lambda_B^*\rb^{T-t-1}(O_1)-\tr\lb O_1 \sigma\rb\one\|_\infty\leq
\|\Lambda_B^{T-t-1}-\Lambda_{B,\infty}\|_{1\to1}.
\end{align*}
We conclude that  in this case, $\delta(l,r)$ can be bounded using mixing time techniques~\cite{Burgarth2013,Temme2010,Reeb2011,Bardet2017,Muller-Hermes2018}. But note that these might provide a too pessimistic bound on $\delta(l,r)$, as they do not take into account the radius of the support $r$. 

Although we made the restrictive assumption that all $\mathcal{U}_t$ are the same, it is straightforward to adapt the arguments above to the case where they are different. This, however, implies that the sequence of quantum channels of interest is not homogeneous in time. It is, in general, not known how to estimate the convergence or even certify convergence for a non-homogeneous sequence. One important exception is when the quantum channels change adiabatically in time~\cite{Hanson2017}.
Moreover, the results of Refs.~\cite{Gonzalez-Guillen2018,1906.11682} seem to indicate that we should expect an exponential decay of the mixing rate function for generic local circuits of logarithmic depth in the number of qubits, but we leave this investigation for future work.
Finally, we note that it is straightforward to adapt our results to the case in which the the unitary channel $\mathcal{U}_t$ depends on a classical random variable. That is, we apply some quantum channel $\mathcal{T}_t$ which is a convex combination of unitaries respecting the locality. This leads to a richer variety of evolutions that can be implemented (see e.g.~\cite{Iten2016}) and can be used to ensure rapid mixing.

\subsection{Correlation length of the produced states}\label{sec:corrlength}
Here we discuss how the mixing rate function $\delta(t,r)$ and the geometry of the interaction scheme can be used to bound the correlations present in the state produced.
We measure the correlations in the state in terms of the covariance, which we introduce below. 
\begin{definition}[Covariance]
Let $E,F$ be observables with disjoint support in $G_T$. Their covariance with respect to a state $\rho$, $\operatorname{cov}_\rho(E,F)$, is defined as:
\begin{align*}
\operatorname{cov}_\rho(E,F)=\tr\lb \rho E\otimes F\rb-\tr\lb\rho E\rb \tr\lb\rho F\rb.
\end{align*}
\end{definition}
We then have:

\begin{prop*}[Correlations of the state]\label{cor:corrlength}
Let $E$ and $F$ be observables whose support is disjoint and contained in a ball of radius $r$ and $\rho=\Phi_{[0,T]}(\rho_0\otimes \rho_B)$. Moreover, let $t_0$ be the largest $t$ s.t. $E_{t}$ and $F_{t}$ have supports that intersect. Then
\begin{align*}
\left|\operatorname{cov}_\rho(E,F)\right|\leq 6\delta(t_0,r)\|E\|_\infty\|F\|_\infty.
\end{align*}
\end{prop*}
\begin{proof}
Note that for $t>t_0$ the supports of $E_t$ and $F_t$ are disjoint by definition, that is, $\Phi_{[t,T]}^*(E\otimes F)$ are still product observables. By the definition of the mixing rate, there are constants $c_E$ and $c_F$ such that:
\begin{align*}
\Phi_{[t,T]}^*(E\otimes F)=\lb c_E\one+\delta(t,r)E'_t\rb\otimes\lb c_F\one+\delta(t,r)F'_t\rb.
\end{align*}
Here $E'$ is an observable satisfying $\|E'_t\|_\infty\leq\|E\|_\infty$ and whose support is contained in the support of $E_t$. Analogous properties apply to $F'_t$. Moreover, note that $c_E\leq \|E\|_\infty$.
Defining 
\begin{align*}
\tilde{C}=\Phi_{[t,T]}^*( \lb c_E\one\otimes \delta(t,r)F'_t\rb+\lb\delta(t,r)E'_t\otimes c_F\one\rb+
\delta(t,r)E'_t\otimes \delta(t,r)F'_t)
\end{align*}
we have that
\begin{align*}
\tr\lb \rho E\otimes F\rb=&\tr\lb \rho_0\otimes \rho_B \Phi_{[0,T]}^*(E\otimes F)\rb
=c_Ec_F+\tr\lb \tilde{C}\rho_0\otimes\rho_B\rb.
\end{align*}
An application of the triangle inequality yields $\|\tilde{C}\|_\infty\leq 3\delta(t,r)\|E\|_{\infty}\|F\|_{\infty}$, from which we conclude
\begin{align}\label{equ:approxcorr}
\left|\tr\lb \rho E\otimes F\rb-c_E c_F\right|\leq 3\delta(t,r)\|E\|_\infty\|F\|_\infty.
\end{align}
A similar computation yields that 
\begin{align*}
\Phi_{[t,T]}^*(E\otimes \one)=
\lb c_E+\delta(t,r)E'_t\rb\otimes \one,\quad
\Phi_{[t,T]}^*(\one\otimes F)=
\one \otimes \lb c_F+\delta(t,r)F'_t\rb.
\end{align*}
We, therefore, have that
\begin{align}\label{equ:indepexpect}
\tr\lb\rho E\rb=c_E+\tr\lb \rho_0\otimes \rho_B\tilde{C}_E\rb,\quad
\tr\lb\rho F\rb=c_F+\tr\lb \rho_0\otimes\rho_B\tilde{C}_F\rb,
\end{align}
where $\tilde{C}_E=\delta(t,r)\Phi_{[t,T]}^*\lb E'_t\rb$ and $\tilde{C}_F$ is defined analogously. From~\eqref{equ:indepexpect} we conclude that:
\begin{align*}
|\tr\lb\rho E\rb \tr\lb\rho F\rb-c_E c_F|\leq 3\delta(t,r)\|E\|_\infty\|F\|_\infty.
\end{align*}
Combining the last inequality with~\eqref{equ:approxcorr} we finally have that:
\begin{align*}
\left|\tr\lb\rho E\rb \tr\lb\rho F\rb-\tr\lb \rho E\otimes F\rb \right|\leq
|\tr\lb\rho E\rb \tr\lb\rho F\rb-c_E c_F|+|\tr\lb \rho E\otimes F\rb-c_E c_F|\leq 
6\delta(t,r)\|E\|_\infty\|F\|_\infty.
\end{align*}
\end{proof}

\subsection{Connection to the results of Kim et al}\label{app:resultskim}
First, we briefly review our assumptions on the noise in the implementation, which are closely related to that of Kim et al.~\cite{kim2017noise,kimswingle}. Like them, we assume that noisy versions $\mathcal{N}_U$ of the required two qubit gates $U$ are implemented, which satisfy:
\begin{align}\label{equ:noisyunitary1}
\|\mathcal{U}-\mathcal{N}_U\|_{\diamond}\leq\epsilon_U.
\end{align}
and the noise acts on the same qubits as $U$.
Here $\mathcal{U}$ is just the quantum channel that corresponds to conjugation with $U$ and $\|\cdot\|_{\diamond}$ is the diamond norm.
Recall that the diamond norm is defined as 
\begin{align*}
\|\Lambda\|_{\diamond}=\sup\limits_{X\in \M_n\otimes \M_n}\frac{\|\Lambda\otimes \textrm{id}\lb X\rb\|_1}{\|X\|_1}
\end{align*}
for a linear operator $\Lambda:\M_n\to\M_n$ and $\|\cdot\|_1$ the trace norm. The diamond norm is a natural way of quantifying the noise in our setting as it also allows us to estimate its effect on systems other than the one the unitary is acting on. However, it should be noted that as all unitaries considered in this work only act nontrivially on two qubits, the diamond norm can differ by at most a factor of $4$ from $\|\cdot\|_{1\to1}$. That is,
\begin{align*}
    \epsilon_U\leq 4\|\mathcal{U}-\mathcal{N}_U\|_{1\to 1}.
\end{align*}
We also assume that the initial state preparation is noisy. This can be modelled similarly by assuming further that all qubits are initialized in a state that is $\epsilon_P$ in trace distance to the ideal one.
Let us now connect the mixing rate function of circuits to stability bounds of noisy implementations, which will allow us to recover~\cite[Theorem 2]{kim2017noise} in our language.
\begin{cor*}[Stability of noisy implementation]\label{thm:generalizedkim}
Let 
\begin{align*}
\rho=\tr_B\left[ \Phi_{[0,T]} \lb\rho_0\otimes \rho_B\rb\right]
\end{align*}
and $\tilde{\rho}$ be the quantum state obtained by replacing every two qubit unitary in $\Phi_t$ by a noisy counterpart satisfying~\eqref{equ:noisyunitary1} and every qubit initialized up to a preparation error of $\epsilon_P$. 
Moreover, let $O$ be an observable supported on a ball of radius $r$ and $\|O\|_\infty\leq 1$. Then for all $0\leq t\leq T$:
\begin{align}\label{equ:stabilityboundkim}
|\tr\lb O\lb \rho-\tilde{\rho}\rb\rb|\leq \delta(t,r)+\sum_{k=t+1}^{T}\delta(k,r)\left[\epsilon_U\lb N_U(k,r)-N_U(k+1,r)\rb+\epsilon_P\lb N_Q(k,r)-N_Q(k+1,r)\rb\right].
\end{align}
\end{cor*}
\begin{proof}
Let $\tilde{\Phi}_t$ be the noisy counterpart of $\Phi_t$. 
As in~\cite[Theorem 2]{kim2017noise}, we now consider the decomposition
\begin{align*}
\Phi_{[0,T]}^*-\tilde{\Phi}_{[0,T]}^*=
\lb\Phi_{[0,t-1]}^*-\tilde{\Phi}_{[0,t-1]}^*\rb\circ \Phi_{[0,t]}^*+
\sum\limits_{k=t}^{T}\tilde{\Phi}_{[0,k-1]}^*\circ\lb \Phi_{k}^* -\tilde{\Phi}_{k}^*\rb\circ \Phi_{[k+1,T]}^*,
\end{align*}
with the convention that $\Phi_{[-1,0]}^*,\Phi_{[T+1,T]}^*$ are the identity.
Let us first estimate the error from the sum by estimating each summand.
First, note that, as before, we have:
\begin{align*}
    \Phi_{[k+1,T]}^*(O)=\delta(k+1,r)A_{k+1}+c_{k+1}\one,
\end{align*}
where once again we have $\|A_{k+1}\|_{\infty}\leq\|O\|_{\infty}$ with the same support as $O_{k+1}$ and $c_{k+1}$ is some constant. Moreover, $\lb \Phi_{k}^* -\tilde{\Phi}_{k}^*\rb$ will map the identity to $0$. Thus,
\begin{align}\label{equ:preliminarycontraction}
\|\tilde{\Phi}_{[0,k-1]}^*\circ\lb \Phi_{k}^*-\tilde{\Phi}_{k}^*\rb\circ \Phi_{[k+1,T]}^*\lb O\rb\|_\infty=
\delta(k+1,r)\|\tilde{\Phi}_{[0,k-1]}^*\circ\lb \Phi_{k}^*-\tilde{\Phi}_{k}^*\rb\lb A_{k+1}\rb\|_\infty
\end{align}
As we assumed that the noise is local, that is, it acts on the same qubits as the two-qubit gate~\footnote{it is possible to treat the case in which the noise acts in a constant neighbourhood of the qubits similarly, but we will not discuss this scenario in order not to overcomplicate the presentation.}
the action of $\tilde{\Phi}_{k}$ and $\Phi_{k}$ will be identical outside the support of $A_{k+1}$. This is because both will just map the identity to the identity outside the support. This implies that only the unitary gates in the past causal cone of the observable contribute to the error and each one by $\epsilon_U$. 
A similar argument holds for the qubit initialization errors, as only erroneous initialization on the past causal cone contribute to the error.
As there are at most $N_U(t-1,r)-N_U(t,r)$ new unitaries at iteration $t-1$ and at most $N_Q(t-1,r)-N_Q(t,r)$ new qubits, we conclude that:
\begin{align}\label{equ:errorsthatcontribute}
\|\tilde{\Phi}_{[0,k-1]}^*\circ\lb \Phi_{k}^*-\tilde{\Phi}_{k}^*\rb\lb A_{k+1}\rb\|_\infty\leq \epsilon_U\lb N_U(k+1,r)-N_U(k,r)\rb+\epsilon_P\lb N_Q(k+1,r)-N_Q(k,r)\rb
\end{align}

Thus, combining~\eqref{equ:preliminarycontraction} and~\eqref{equ:errorsthatcontribute} yields:
\begin{align*}
&\|\sum\limits_{k=t}^{T}\tilde{\Phi}_{[0,k-1]}^*\circ\lb \Phi_{k}^*-\tilde{\Phi}_{k}^*\rb\circ \Phi_{[k+1,T]}^*\lb O\rb\|_\infty\leq
\sum\limits_{k=t}^{T}\delta(k+1,r)\|\tilde{\Phi}_{[0,k-1]}^*\circ\lb \Phi_{k}^*-\tilde{\Phi}_{k}^*\rb \lb A_{k+1}\rb\|_\infty\\
& \leq\sum_{k=t}^{T}\delta(k+1,r)\left[\epsilon_U\lb N_U(k,r)-N_U(k+1,r)\rb+\epsilon_P\lb N_Q(k,r)-N_Q(k+1,r)\rb\right]
\end{align*}
Now, by the definition of the mixing rate function there exists an observable $A$ such that
\begin{align*}
\Phi_{[k,T]}(O)=c \one+\delta(k,r)A
\end{align*}
with $\|A\|_\infty\leq1$. Thus, we see that 
\begin{align*}
\lb\Phi_{[0,k-1]}^*-\tilde{\Phi}_{[0,k-1]}^*\rb\circ \Phi_{[k,T]}(O)=
\delta(k,r)\lb\Phi_{[0,k-1]}^*-\tilde{\Phi}_{[0,k-1]}^*\rb\circ \Phi_{[k,T]}^*(A),
\end{align*}
as the identity is in the kernel of $\Phi_{[0,k-1]}^*-\tilde{\Phi}_{[0,k-1]}^*$. We conclude that 
\begin{align*}
\|\Phi_{[0,T]}^*-\tilde{\Phi}_{[0,T]}^*(O)\|_\infty\leq \delta(t,r)+\sum_{k=t}^{T}\delta(k,r)\left[\epsilon_U\lb N_U(k,r)-N_U(k+1,r)\rb+\epsilon_P\lb N_Q(k,r)-N_Q(k+1,r)\rb\right],
\end{align*}
from which the claim follows.

\end{proof}
The stability results of Refs.~\cite{kim2017noise,kimswingle} are captured by this corollary. For instance, the main result of Ref.~\cite{kim2017noise} follows from assuming that there exist constants $r_0,c,k,\alpha,\Delta\geq 0$ independent of system size such that for all $r\leq r_0$:
\begin{align*}
\delta(t,r)=cr^{\alpha}e^{-\gamma (T-t)}+\Delta.
\end{align*}
Optimizing
\begin{align*}
t_0=T-\frac{1}{\gamma}\log\lb\frac{\epsilon}{Dr^{\alpha}c}\rb^2
\end{align*}
suffices to guarantee an estimate up to $\cO\lb D^2\epsilon\log(\epsilon^{-1})^2+\Delta\rb$, as in~\cite{kim2017noise}. 
By comparing Corollary~\ref{thm:generalizedkim} with our main theorem (see article), we see that this stability comes from the fact that the assumptions on $\delta(t,r)$ imply that there is an "effective" circuit of constant size underlying the computation. 
Moreover, each iteration of the evolution can only change the expectation value by an amount that decreases with time. 

This is well illustrated when we compare the bound in Eq.~\eqref{equ:stabilityboundkim} and the one we obtained with our main result, reproduced in the supplementary material in Eq.~\eqref{equ:ourstabilitybound}. Note that the two bounds only differ by a factor of $\delta(t,r)$. This difference has a clear interpretation in light of the discussion above: in our result we allowed for an arbitrary initial state $\tilde{\rho}$ when implementing the past causal cone from iteration $t$ to $T$ , while above the state at iteration $t$ is given by $\Phi_{[0,t-1]}(\rho_0\otimes \rho_B)$ in the noiseless version. With the previous discussion in mind, we see that \emph{any} change to the state produced from iteration $0$ to $t$ can only change the expectation value by $\delta(t,r)$, which explains the extra $\delta(t,r)$ factor.
It is also important to note that the connection between rapid mixing under local evolutions and stability of the expectation values of local observables was established in~\cite{Lucia2015,Cubitt2015}, where the authors show similar results for time evolutions in continuous time.

Finally, it should be noted that our approach also requires a bound on $\delta(t,r)$ to ensure that the energy inferred from the smaller patches corresponds to a physical quantum state. This is not the case if the whole circuit or causal cone preparing the many-body state is implemented.
\subsection{Certifying mixing}\label{sec:mixing}
A close look at the proof of the main theorem shows that $\delta(t,r)$ provides a worst-case estimate for how fast the expectation values stabilize. If we are only interested in estimating the expectation value of a given observable $O$, we see that
\begin{align*}
    \inf\limits_{c\in \R}\|O_t-c\one\|_\infty
\end{align*}
gives an upper bound on the error we obtain when we estimate $\tr(\rho O)$ by only implementing the circuit from iteration $t$ to $T$. Thus, it is not necessary to bound the mixing rate for arbitrary observables, which is expected to be hard in general. E.g. the results of~\cite{Bookatz2013} show that it is QMA-hard to determine the spectral gap~\cite{Temme2010} of certain quantum channels, which is a central quantity in determining the mixing time of quantum channels.
We will therefore focus on bounding the mixing for a given observable $O$. We will show that in case $\operatorname{supp} O_t$ is small it is possible to bound $\|O_t-c\one\|_\infty$ on a quantum computer. 

As can be seen in the proof of the main theorem, if $\delta(t,r)$ is small, then the output of the circuit is essentially independent of the initial state. Thus, it should be expected that the dependence of the expectation value of an observable $O$ on the initial state gives an estimate on the mixing time. Indeed, if we draw a state $\sigma_t$ from a state two design~\citep{Ambainis2007} on the support of $O_t$ and define the random variable $X_t=\tr\lb \sigma_t O_t\rb$, then:
\begin{align}\label{equ:variance}
&\lb 2^{n_t}\lb \mathbb{E}(X_t^2)\lb 2^{n_t}+1\rb-2\mathbb{E}\lb X_t\rb^2\rb\rb ^{\frac{1}{2}}\geq 
\|O_t-\tr\lb \Phi^{*}_{[t,T]}\lb O\rb\rb\frac{\one}{2^{n_t}}\|_\infty.
\end{align}
Here $n_t$ is the number of qubits on the support of $O_t$.
As it is possible to generate a two state design using $\cO(n_t\log^2(n_t))$ gates~\cite{Cleve2001}, equation~\eqref{equ:variance} gives a protocol to measure how far each local observable is from stabilizing as long as $n_t$ is small by estimating the first and second moments of $X_t$. 
This protocol applies to interaction schemes for which the support  of observables has a bounded radius, like DMERA, otherwise the scaling in $n_t$ is prohibitively large. 

We now discuss to derive~\eqref{equ:variance} and its consequences in more detail.
We start by recalling the definition of a quantum state design~\cite{Ambainis2007}:
\begin{definition}[State design]
A distribution $\mu$ over the set of $d$ dimensional quantum states is called a $k-$state design for some $k>0$ if
\begin{align*}
\int(|\psi\rangle\langle\psi|)^{\otimes k} d\mu=\int(|\psi\rangle\langle\psi|)^{\otimes k} d \mu_U,
\end{align*}
where $\mu_U$ is the (normalized) uniform measure on the set of pure quantum states.
\end{definition}
That is, these states have the same first $k$ moments as the uniform distribution on the set of pure states. Let us now compute some relevant moments of the random quantum states:
Let $\ket{\psi}$ be drawn from the uniform distribution of $d-$ dimensional pure quantum states and $O$ be an observable. Moreover, define the random variable $X=\tr\lb \ketbra{\psi}{\psi} O\rb$.
Then:
\begin{align}\label{equ:firstmoments}
\mathbb{E}(X)=\frac{\tr\lb O\rb}{d},\quad\mathbb{E}(X^2)=\frac{1}{d(d+1)}\lb \tr(O^2)+\tr(O)^2\rb.
\end{align}
This can be derived by e.g. noting that $\ketbra{\psi}{\psi}$ has the same distribution as $U\ketbra{0}{0}U^\dagger$, where $U$ is a Haar random unitary. 
A simple application of the Weingarten calculus for the moments of the Haar measure on the unitary group~\cite{Collins2006,1902.08539} yields the result.
We are now ready to prove equation~\eqref{equ:variance}, which we restate as a lemma for the reader's convenience:
\begin{lemma*}[Checking mixing]
Let $O$ be an observable and $n_t$ be the number of qubits in the support of $O_t$. Moreover, let $\sigma_t$ be drawn from a state $2-$design on the support of $O_t$ and denote by $X_t$ the random variable $X_t=\tr\lb \Phi_{[t,T]}\lb\sigma_t\rb O\rb$. Then
\begin{align*}
\lb 2^{n_t}\lb \mathbb{E}(X^2)\lb 2^{n_t}+1\rb-2\mathbb{E}\lb X\rb^2\rb\rb ^{\frac{1}{2}}\geq 
\|\Phi^{*}_{[t,T]}\lb O\rb-\tr\lb \Phi^{*}_{[t,T]}\lb O\rb\rb\frac{\one}{2^{n_t}}\|_\infty.
\end{align*}
\end{lemma*}
\begin{proof}
Note that 
\begin{align*}
\left\| \Phi^{*}_{[t,T]}\lb O\rb-\tr\lb \Phi^{*}_{[t,T]}\lb O\rb\rb\frac{\one}{2^{n_t}}\right\|_F^2=
\tr \lb \Phi^{*}_{[t,T]}\lb O\rb^2\rb-2^{-n_t}\tr\lb \Phi^{*}_{[t,T]}\lb O\rb\rb^2.
\end{align*}
Here $\|\cdot\|_F$ is the Frobenius norm.
It follows from~\eqref{equ:firstmoments} that
\begin{align*}
2^{n_t}\lb \mathbb{E}(X^2)\lb 2^{n_t}+1\rb-2\mathbb{E}\lb X\rb^2\rb=
\left\| \Phi^{*}_{[t,T]}\lb O\rb-\tr\lb \Phi^{*}_{[t,T]}\lb O\rb\rb\frac{\one}{2^{n_t}}\right\|_F^2
\end{align*}
if we draw $\sigma$ from the uniform distribution on states. But it is clear that the expression only depends on the second and first moments of the random variable. Thus, a state $2-$design satisfies the same properties.
The claim then follows from the fact that $\|\cdot\|_F\geq\|\cdot\|_\infty$.
\end{proof}
We note that in case the quantum channel has a spectral gap, then 
\begin{align*}
\left\| \Phi^{*}_{[t,T]}\lb O\rb-\tr\lb \Phi^{*}_{[t,T]}\lb O\rb\rb\frac{\one}{2^{n_t}}\right\|_F^2
\end{align*}
decays exponentially with $(T-t)$~\cite{Temme2010}.
As a quantum state two design of $n$ qubits can be generated with a circuit consisting of $\cO(n\log^2(n))$ two-qubit gates~\cite{Cleve2001}.
Moreover, for some $n$ there are simplified constructions of state $2$-designs that only require a circuit of linear depth and Hadamard and controlled phase gates~\cite{Seyfarth_2011}. These are based on the fact any set of maximal mutually orthogonal bases also gives a two-design~\cite{Klappenecker_2005}.

Moreover, it is important to take into account the effect of noise in the evolution. As the proof above did not take into account any special property of the quantum channel itself, we can replace $ \Phi^{*}_{[t,T]}$ by the noisy evolution $ \tilde{\Phi}^{*}_{[t,T]}$ and the statement still holds. Thus, assuming that we implement the noisy version instead and can prepare the states in the $2$-design perfectly, the protocol above still measures how mixed the outputs of  the noisy quantum computer.

Thus, it only remains to estimate the effects of noise in the preparation procedure of the $2$-design required to estimate the moments of $X_t$.  Unfortunately, the $2^{n_t}$ pre-factor in the inequality above implies that the precision in the preparation of the $2-$design and number of samples required to check mixing is infeasible whenever the support of $O_t$ is large, as both scale exponentially with $n_t$.

\bibliography{biblio}